\documentclass[10pt,twocolumn,letterpaper]{article}

\usepackage{wacv}
\usepackage{times}
\usepackage{epsfig}
\usepackage{graphicx}
\usepackage{amsmath}
\usepackage{amssymb}
\usepackage{booktabs}
\PassOptionsToPackage{warn}{textcomp}  % to address font issues with \textrightarrow
\usepackage{textcomp}                  % use better special symbols
\usepackage{mathptmx}                  % use matching math font
\usepackage{times}                     % we use Times as the main font
\usepackage{subcaption}
\usepackage{authblk}

\def\wacvPaperID{1828} % Enter the WACV Paper ID here

\wacvapplicationstrack % Uncomment this line if you are submitting to the Applications Track.

\wacvfinalcopy % *** Uncomment this line for the final submission

\ifwacvfinal
\usepackage[breaklinks=true,bookmarks=false]{hyperref}
\else
\usepackage[pagebackref=true,breaklinks=true,colorlinks,bookmarks=false]{hyperref}
\fi

\begin{document}

%%%%%%%%% TITLE
% \title{DIAGNOSIS: A Diffusion Probabilistic Based Model for Genotypes Guided Synthesis of Histopathology Images}

\title{A Morphology Focused Diffusion Probabilistic Model for Synthesis of Histopathology Images}

\newcommand\CoAuthorMark{\footnotemark[\arabic{footnote}]} % get the current value
% \author[2,3]{Darth Sidious\protect\CoAuthorMark
%   \thanks{Corresponding author}}
 
\author[1]{Puria Azadi Moghadam\protect\CoAuthorMark\thanks{Equal Contribution}}
\author[2]{Sanne Van Dalen \protect\CoAuthorMark}
\author[3]{Karina C. Martin}
\author[4,5]{Jochen Lennerz}
\author[3,6]{Stephen Yip}
\author[7]{Hossein Farahani}
\author[3,7]{Ali Bashashati}
\affil[1]{ Department of Electrical and Computer Engineering, University of British
Columbia, Canada}
\affil[2]{Eindhoven University of Technology}
\affil[3]{ Department of Pathology and Laboratory Medicine, University of British
Columbia, Canada}

\affil[4]{Center for Integrated Diagnostics, Department of Pathology, Massachusetts General Hospital, USA}
\affil[5]{Department of Pathology, Harvard Medical School, USA}
\affil[6]{Molecular Oncology, BC Cancer Agency, Canada}
\affil[7]{School of Biomedical Engineering, University of British Columbia, Canada}

% \author{Puria Azadi Moghadam, \\
% Department of Electrical and Computer Engineering,\\
% University of British Columbia, \\
% Vancouver, Canada\\
% {\tt\small puria.azadi@ubc.ca}
% \and

\maketitle
\thispagestyle{empty}

%%%%%%%%% ABSTRACT
\begin{abstract}
%   Low-grade gliomas are classified into sub types on the basis of histopathology images. However, 
Visual microscopic study of diseased tissue by pathologists has been the cornerstone for cancer diagnosis and prognostication for more than a century.  
Recently, deep learning methods have made significant advances in the analysis and classification of tissue images. However, there has been limited work on the utility of such models in generating histopathology images. 
These synthetic images have several applications in pathology including utilities in education, proficiency testing, privacy, and data sharing. 
Recently, diffusion probabilistic models were introduced to generate high quality images. 
Here, for the first time, we investigate the potential use of such models along with prioritized morphology weighting and color normalization to synthesize high quality histopathology images of brain cancer.
Our detailed results show that diffusion probabilistic models are capable of synthesizing a wide range of histopathology images and have superior performance compared to generative adversarial networks.
\end{abstract}

%%%%%%%%% BODY TEXT
% \section{Introduction}
% \section{Introduction}


\section{Introduction}\label{sec:introduction}
% \textbf{Histo, digital Histophathology and staining }
% Histopathology is a diagnostic discipline founded on the visual interpretation of cellular biology captured in images.

% Whole slide imaging, also known as virtual microscopy, refers to scanning a complete microscope slide and creating a single high-resolution digital file. Automated digital pathology scanners, it is possible to capture an entire glass slide, under bright field or fluorescent conditions, at a magnification comparable to a microscope.

% Histological staining is a series of technique processes undertaken in the preparation of sample tissues by staining using histological stains to aid in the microscope study (Anderson, 2011). Staining is used to highlight important features of the tissue as well as to enhance the tissue contrast.

\begin{figure}[h]
    \includegraphics[scale=.8]{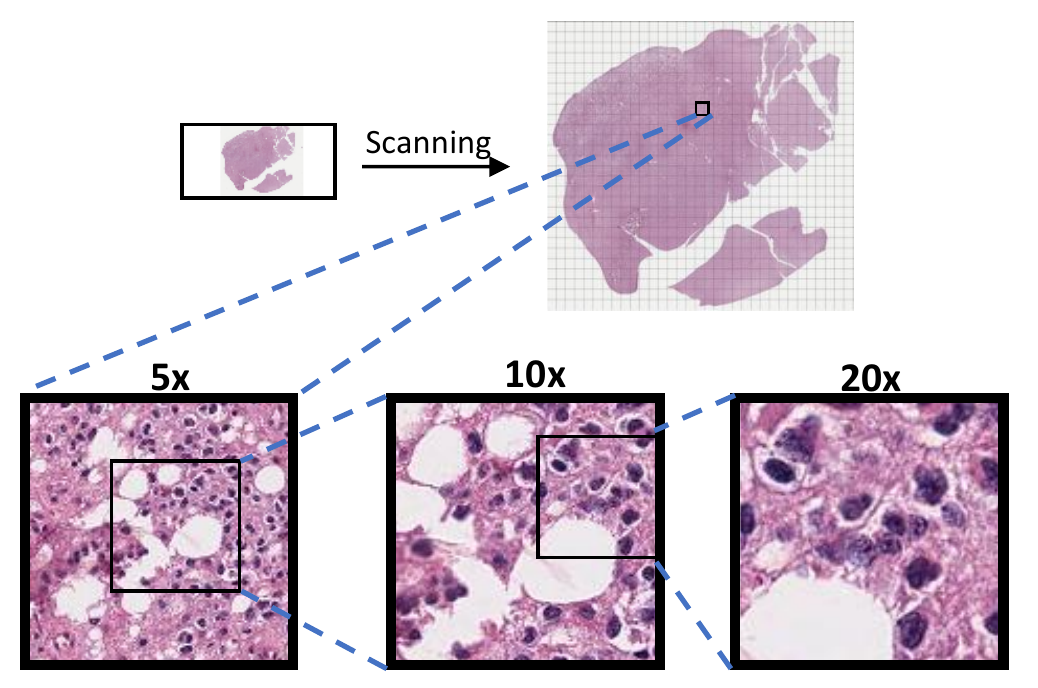}
    \caption{An example of whole slide image and 3 sampled patches at different magnifications.}
    \label{fig:HistoIntroduction}
\end{figure}

Histopathology is a diagnostic science that relies on the visual examination of cellular and tissue characteristics in magnified tissue slides\cite{WhatHisto}. Recently, high-throughput digital pathology scanners have been developed that can provide gigapixel high-resolution images($\sim 100K\times100K$ pixels) of microscope slides at objective magnifications of up to 40$\times$. Furthermore, histological staining of tissues with various stains (e.g., hematoxylin and eosin, silver nitrate, carmine, hematin, etc.) is used to emphasise the properties of the tissues and improve their contrast for examination ~\cite{stain}. %These steps among other minor preprocessings are utilized in histopathology for decision making and refers to whole slide imaging(WSI). 
\autoref{fig:HistoIntroduction} shows a sample of gigapixel pathology images.

% XXX Create the figure for digital histopath XXX

% \textbf{Be careful for plagerism}
 
% The tissue sample is stained to increase the contrast between different structures.
The most commonly used stain material is Hematoxylin and Eosin (H\&E), which stains the nucleic acid within the cell nuclei with purplish blue and extracellular matrix and cytoplasm by pink color. Afterwards, pathologists examine the cytological and tissue characteristics of the sample for cancer diagnosis and staging. The histopathological diagnosis of cancer is time-consuming and is prone to subjective differences\cite{process}, 
% The process consists of manually examining both small and big resolutions of images. 
% In addition, a detailed and extensive analysis is required since the difference between the sub types are hard to find. 
% Discuss => 
%Pathologists must manually examine images, which is difficult and time-consuming for the majority of them. 
as it is heavily reliant on pathologists' experiences and prior exposure to various histological variants (i.e., subtypes). However, some of these variants are rare, and pathologists do not get the opportunity to examine them during their training. In addition, there is a scarcity of samples for certain subtypes in various parts around the world. 
% , and there is a need to have a less biased process in training pathologists.

% \textbf{Copied from Ali's paper. paraphrase it}
Following the success of deep learning models in the past decade in other applications, various models have been developed in recent years to improve the quality of histopathology diagnosis and assist pathologists in the decision making process. However, majority of the research efforts have been on developing discriminative models with more emphasis on classification tasks ~\cite{Histo-DL_-survey}. 
%The majority of these methods are focused on classification. 
Wang et al.~\cite{CellDetection}, for example, employed an accelerated deep convolution neural network for both cell identification and classification. Mathur et al.~\cite{WACV_histo} developed a multi-gaze attention network with multiheaded self-attention for renal biopsy image categorization tasks. 
Segmentation is another frequent application of deep models in histopathology. The authors in~\cite{MultiRes} presented their multi-resolution UNet-based approach for detecting breast cancer metastasis. Another example is the use of a deep learning algorithm for the prediction of patients' outcomes\cite{predict-outcomes}. 

% There is a considerable gap between the mentioned models and generative models in pathology. In the recent years, synthesizing theqniques hve imporved but not in histopathology.
% Although these examples show promising results, they all used discriminative machine learning methods, which map data to a class label, while the use of generative modelling in pathology is still relatively unexplored. 

% \textbf{Copied from Ali's paper. paraphrase it}
Compared to discriminative models, utilizing generative models in pathology is still in its infancy. Generative models can be used to create synthetic images and have numerous potential applications in pathology and can entail discovering patterns and regularities that may give more information about distinctions across subtypes.

They have the potential to improve the educational, assessment, security, and generalization aspects of the histology field.
Moreover, deep generative models can be utilized to synthesize histology images of rare cancer subtypes for training pathologists and proficiency testing ~\cite{Oversampling}. 
In addition, these images can tackle the privacy concerns of sharing pathology images, which is one of the emerging challenges with patient privacy and data protection\cite{privacy}. In comparison to normal images, several factors, such as time, labor, and economic costs, make having fully labeled medical images more difficult~\cite{GANinMedical}, causing the model to suffer from overfitting. Synthesized histology images can be used to augment the  datasets and improve the performance of trained models.

Generative Adversarial Networks (GAN) introduced by Goodfellow et al.\cite{gans-ian} and its later variants are currently the most common models for generating synthetic images.
However, GANs have several drawbacks, such as mode collapse and difficulties in training. %Thus, there is a need for a new generative model which outperforms GANs.

Starting from Ho et al.\cite{ho-diffusion} several studies showed that diffusion probabilistic models can generate high-fidelity images comparable to those generated by GANs~\cite{diffusion-nichol, beat-gans, p2-weighting}. Diffusion probabilistic models offer several desirable properties for image synthesis, such as stable training, easy model scaling, and good distribution coverage. However, their performance has not been explored for histopathology images.

% AS there is no mode collapse, it is better for generating images due to fact that there are rare classes for generating.

% \textbf{summarize this like a real paper}
% In this paper, diffusion probabilistic models are implemented to generate synthetic histopathology images of low-grade gliomas. Morover, conditional synthesis is used to create histopathology images of three low-grade gliomas sub types, namely the IDH wildtype, IDH mutation with 1p/19q codeletion, and IDH mutation without 1p/19q codeletion. The quality of the synthetic images generated by different diffusion probabilistic models are compared using visual inspection. 
% histo features and patterns are diff and  rare genotypes make it hard with mode collapse
The goal of this paper is to explore the utility of diffusion
probabilistic models for synthesizing histopathology images and comparing these models with state-of-the-art for quality of the generated images and lack of artefacts. 
%Due to highly unbalanced data for some of pathology branches subtypes in the wild and different nature of histology images compared to normal images generating histopathlogy images is a challenging problem.
% since there is highly unbalanced data for some of subtypes in the wild and also due to different nature of histology images compared to normal images.
%This causes significant damages in downstream applications that require synthetic histology images. Also, an accurate diagnosis regarding the tumor is essential for determining the optimal treatment  and weighing up the potential benefits against risk of treatment.
%We propose a replacement to the current generative models in histopathology. The proposed method greatly facilitate the deployment of augmented histopathology datasets for many real-life applications.

The contributions of this paper are as follows:
\begin{itemize}
    
    \item In this paper, for the first time, we propose exploiting diffusion probabilistic models to generate synthetic histopathology images.% of low-grade gliomas(LGG). This model can easily be adapted to other cancer types.
    
    \item We also benefit from color normalization to force our end-to-end model to learn morphological patterns and from perception prioritized weighting (P2)~\cite{p2-weighting}, which aims to prioritize focusing on diffusion stages with more important structural histopathology contents.
    
    % \item In this paper, we propose exploiting diffusion probabilistic models for the first time to generate synthetic histopathology images of low-grade gliomas(LGG) that are associated with different genotypes. We also benefit from color normalization to force the model to learn morphological patterns and from perception prioritized weighting (P2)~\cite{p2-weighting}, which aims to prioritize focusing on noise levels with more important perceptual contents. This model can easily be adapted to other cancer types.
    
%    \item We introduce our unique tumor vs normal area pixel-wise annotations done by a board-certified pathologist or a pathology resident under the supervision of a board-certified pathologist for a large selected cohort of patients from TCGA LGG dataset. This carefully annotated data can be used for training both patch level generative and discriminative histology models covering various applications. We are going to publicly release our code, selected patients in our cohort, and their slides' annotations.
    
    \item We conduct an extensive empirical study using a low grade glioma (LGG) dataset and compare the performance of the proposed generation method against a state-of-the-art study that utilized GANs for histpathology image analysis, using multiple objective and subjective metrics. Our results show that the proposed method outperforms in all metrics, and it produces pathology images that are close to the ground truth. Our datasets and experimental results are described in~\autoref{sec:results}.
    
    \end{itemize}
    
\section{Related Works}\label{sec:related}
\begin{figure*}[h]
    \includegraphics[width=.986\textwidth]{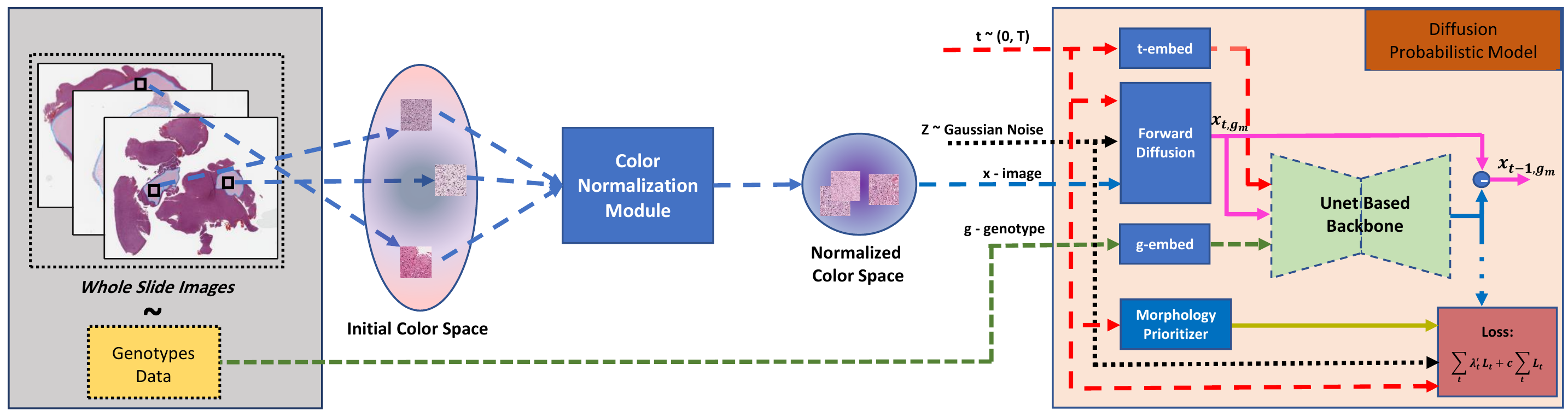}
    \caption{The overview of proposed approach for generating histopathology images with genotypes guidance}
    \label{fig:model-overview}
\end{figure*}

% Talk about GAN and some new GAN including ProGAN. then one other work and levin work. say are GAN based with probelems.

Generating Histopathology images has been getting popular in recent years because of advances in digital pathology imaging and computational infrastructures as well as the introduction of powerful deep generative models that are able to tackle specific difficulties in this domain.

% Zuru, other pro in onenote for classi. the other pro for superres by GAN. Finally levine. seems ProGAN idea is populur in synthesizing. say bad GAn from Vahdat work. 

In 2018, Kapil et al.~\cite{GAN_first} utilized generative learning to enable automated tumor proportion scoring. Zhou et al.~\cite{SuperResGAN} employed a U-net based GAN to augment and generate high-resolution image from low-resolution histology patches.
More recently, the Progressive GAN model has shown considerable potential in generating high quality histopathological images~\cite{paper-ali, xue2021selective}.

% the Levine et al.\cite{paper-ali} showed the potential of GANs to synthesize high resolution pathology images of ten histological types of cancer. 
% Why GANs are bad and diffusions are good for unbalanced and no mode collapse: 

GAN models, on the other hand, suffer from mode collapse and instabilities due to directly producing images from complex latent spaces in a single shot and easily overfitting of their discriminator\cite{vahdat-problem}, which make them unsuitable for generating samples from rare conditions or imbalanced datasets.

% GAN models, on the other hand, suffer from mode collapse, instabilities, and difficulty in training, as well as readily overfitting of the discriminator when just looking at clean genuine images because they directly generate images from a complicated distribution in a single step.

%  However, there are some drawbacks with GANs. One major drawback(cite) of GANs is mode collapse, which means that they may only characterize a few modes of the true distribution. limit the modes. Furthermore, GANs are difficult to train. Thus, there is a need for a new generative model which outperforms GANs. 

However, diffusion probabilistic models were recently introduced and used for several applications. Amit et al.\cite{amit} used a diffusion model for performing image segmentation. Other examples are the use of diffusion probabilistic models for converting text to speech\cite{popov} and generating online handwriting\cite{luhman}. As diffusion models divide the process into several relatively simple denoising diffusion steps and make strong conditioning on input images at each step, they soften the data distribution, which results in models with higher stability. These models also are able to generate more diverse images with better distribution coverage, are less probable to get overfitted, and are easy for  scaling~\cite{vahdat-problem}.

These advantages can significantly improve the histopathology image generation domain as imbalanced datasets with rare subtypes is one of the main issues in this domain. Although diffusion probabilistic models are used within different areas for multiple tasks, they are not exploited to generate synthetic histopathology images yet. Therefore, we hypothesize that such models could generate high quality histopathology images that may address some of the challenges of GAN-based models. 

% Their diversity can be big help for pathology which suffers from rare cases.

%  overfitting discriminator is the other reason.
 
% \textbf{summarize at the end of related works}
% Diffusion probabilistic models offer several desirable properties for image synthesis, such as stable training, easy model scaling, and a good distribution coverage. Starting from Ho et al.\cite{ho-diffusion} several studies showed that diffusion probabilistic models can generate high-fidelity images comparable to those generated by GANs\cite{diffusion-nichol}\cite{beat-gans}\cite{p2-weighting}. 

% AS there is no mode collapse, it is better for generating images due to fact that there are rare classes for generating.

\section{Method}\label{sec:method}

% \section{Methods}

\subsection{Problem Definition}
% \textbf{Summarize next 2 paragraph and be careful for plagerism ---> Move low Grade Glioma to problem definition}

% IDH mutation is highly common in LGG. Furthermore, codeletion of 1p/19q was found in about 60-90 percent of diagnose oligodendroglioma and 30-50 percent of oligoastrocytma, while astrocytomas lack 1p/19q codeletion. Moreover, differences between these genotype features has been associated with the prediction of tumor response and prognosis. For example, the 1p/19q co-deletion has been identified as a significant marker for a better response to treatment and a better prognosis. In addition, the impact of treatments on preserving quality of life and neurocognition is important. Therefore, an accurate diagnosis is important for determining the optimal treatment and weighing up the potential benefits against risk of treatment.

The objective of this paper is to enable the generation of histopathology images that are represented by various morphologic and genomics features. Synthesizing these pathology images is a challenging task compared to typical images in other domains.

Assume $g_{i=1,2,..k}$ indicates the i-th genotype available in nature. For each genotype, there is:
\begin{equation}
    Set_{g_i} = \{images_j\}, j=1,2,...,N \; \& \; image_j \in R^{c\times h \times w},
\end{equation}
which:
\begin{equation}
    genotype(image_j) = g_i,
\end{equation}
The purpose is to have an estimator function $f_{est}$ that: 
\begin{equation}
     f_{est}{ \{g_i, n  \sim Noise  \}  }= image_{g_i, out},
\end{equation}
where:
\begin{equation}
     image_{g_i, out} \in R^{c\times h \times w}, \; \; image_{g_i, out} \sim dist \left ( Set_{g_i} \right ),
\end{equation}

we tackle LGG, which account for the majority of pediatric brain tumors~\cite{LGG_molicular}
% and encompass a wide variety of histopathologies.
and they are classified by combining the histopathological features with genotype features since 2016\cite{who-classification} 
% Common types among this group are astrocytomas, oligodendrogliomas, and mixed oligoastrocytma.
including isocitrate dehydrogenase (IDH) and co-deletion of 1p19q which are the short arm of chromosome 1 and the long arm of chromosome 19\cite{StY1, StY2}. Diagnosis of LGG is done through histopathologic examination of tissue.
% Worth noting that the various LGG's genotypes are not distributed equally in the wild. 
\autoref{fig:model-overview} summarizes our proposed end-to-end solution for generating histology images.

\subsection{Color normalization}
% Cite papers that show color norm improve discriminative since focus on morphological features to classify. then, we plan to use the same techiq for generation.         
% To force it to learn motphological features such as cell shapes and distribution that are used by patho to cure instead of staining differences. since may have slight color differences.

One of the main challenges related to H\&E images is the lack of consistency in the staining due to variances in site-specific staining protocol or digital scanning platforms and methodologies. Color normalization strategies are able to boost histological discriminative models' performance.\cite{boschman-paper}. We propose employing the same strategies to convert the input images to a same color domain in order to derive the diffusion model focus on learning the morphological patterns and other vital pathology aspects such as cell shape, density, and distribution rather than stain differences.

\begin{figure}[tp]
\begin{center}
    \includegraphics[scale=0.74]{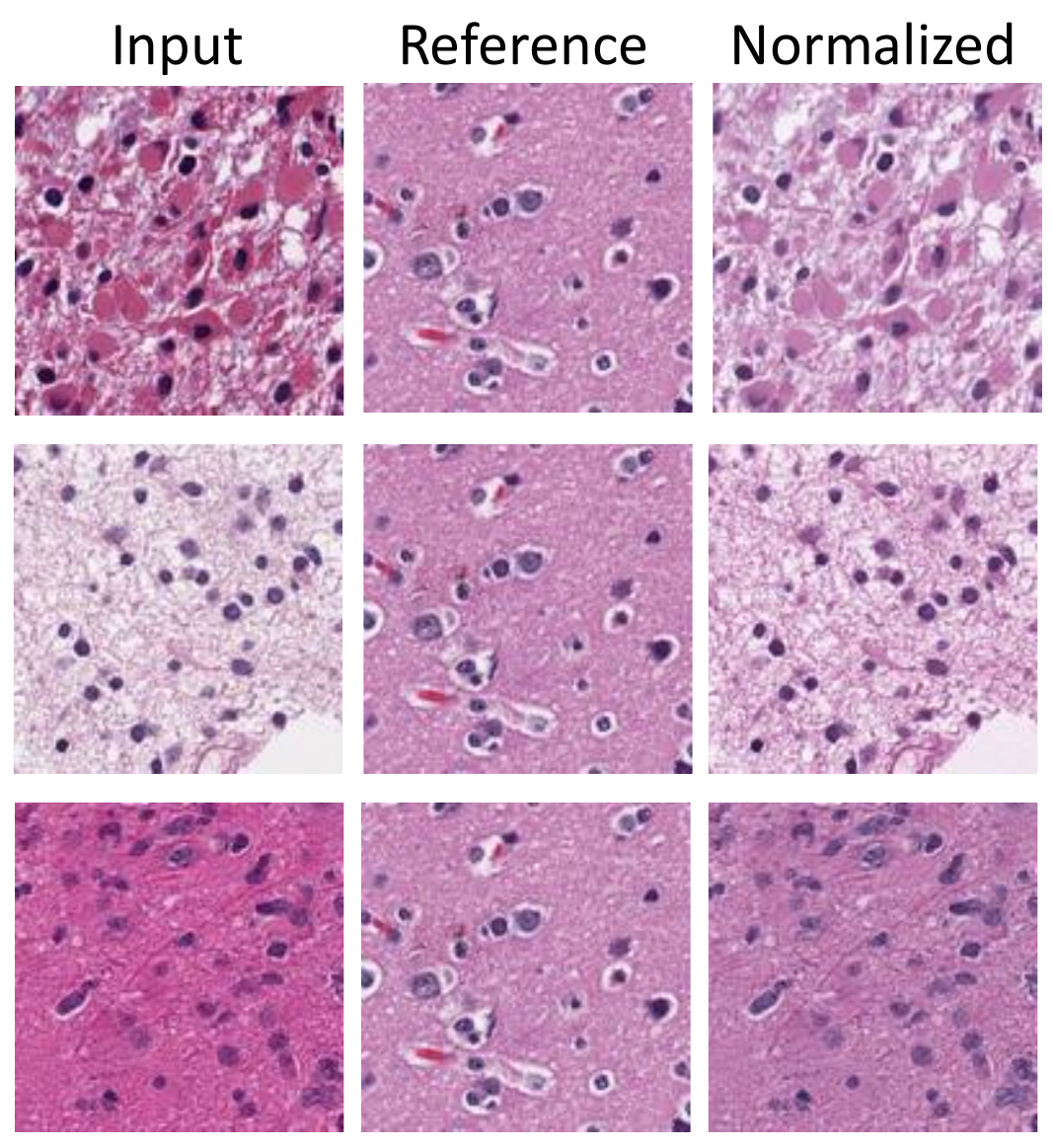}
    \caption{Color Normalization Visualization}
    \label{fig:ColorNorm}
\end{center}
\end{figure}

For color normalization, we used the structure-preserving color normalization scheme introduced by Vahadane et al.\cite{color-vahadane} that transfers source images to the target domain while preserving their own stain concentration.
\autoref{fig:ColorNorm} visualizes the performance of the color normalization method on three extracted patches of histopathology images for a same reference.

\subsection{Diffusion Probabilistic Model}
\begin{figure*}[tp]
    \includegraphics[scale = 0.76]{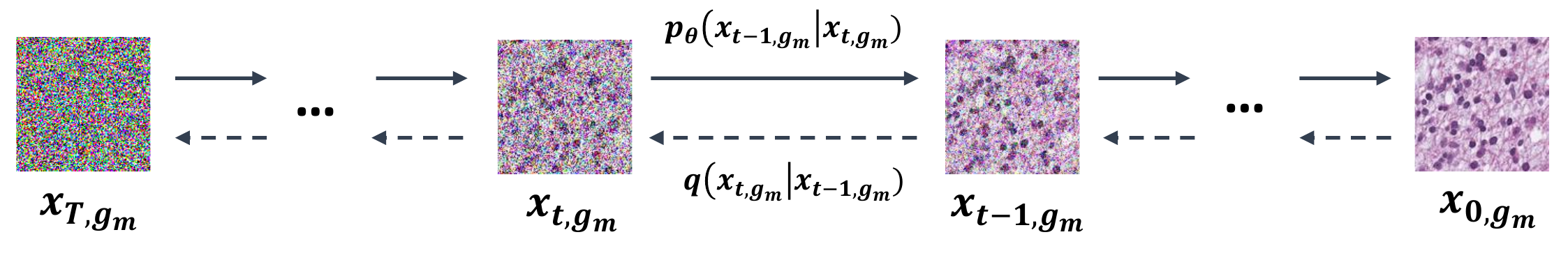}
    \caption{The illustration of the forward(0 to T) and reverse(T to 0) diffusion process}
    \label{fig:model-presentation}
\end{figure*}

% \textbf{transfer all x to tuple (x, genomic)}
% The Pytorch\cite{pytorch} implementation of the diffusion probabilistic model\cite{beat-gans} was used to generate synthetic histopathology images.
The diffusion model can be summarized in two main processes: forward diffusion and parameterized reverse diffusion. 
\autoref{fig:model-presentation} illustrates the two directions of diffusion probabilistic models. The model in the former process gradually generates noisier samples from real data using Gaussian noise kernel. The later process makes the model able to iteratively retrieve data from noise which can be employed to produce synthetic data from random noise.\\

\subsubsection{Forward Diffusion} 
Let $x_{_0, g_m}$ be the real input data from the m-th genotype $g_m$ and $x_{_t, g_m}$ be the noisy images for $g_m$ produced at time $t=1,2,...T$.
% Starting from a image from the input data, $x_0$, the diffusion process gradually adds Gaussian noise with predefined noise scales. $0<\beta_{1},\beta_{2},...,\beta_{T}<1$, indexed by time step $t$. 
% This diffusion process $q$ is a Markov chain and produces latents $x_{_1, g_m}$ through $x_{_T, {g_m}}$ as follows~\cite{ho-diffusion}:
% \begin{equation}
%     q(x_{_1, g_m}, ...., x_{_T, g_m}|x_{_0, g_m}) = \prod_{t=1}^{T} q(x_{_t, g_m}|x_{{_{t-1}, g_m}}),
% \end{equation}
% \begin{equation}
%     q(x_{_t, g_m}|x_{_{t-1}, g_m}) = \mathcal{N}\left (x_{_t, g_m};\left (\sqrt{1-\beta_{t}} \right) x_{_{t-1}, g_m},\beta_{t}\mathit{I}\right ),
% \end{equation}
% and by using reparameterization, 
Latent $x_{_t,{g_m}}$ can be derived directly from $x_{_0, g_m}$ as following~\cite{ho-diffusion}:
\begin{equation}
    x_{_t,g_m} = \sqrt{\alpha_{t}}x_{_0,g_m}+\sqrt{1-\alpha_{t}}\epsilon,
    \label{eq:eq12}
\end{equation}
\let\subsectionautorefname\sectionautorefname
\let\subsubsectionautorefname\sectionautorefname
$0<\beta_{1},\beta_{2},...,\beta_{T}<1$ are fixed noise scales for each time step $t$ and $\alpha_t := \prod_{s=1}^{t}(1-\beta_s)$. Also, the distribution of the$\epsilon$ is as  $\epsilon \sim \mathcal{N}(0,I)$. (Details are in section 1.1 of supplementary material)
\subsubsection{Parametrized Reverse Diffusion}
In order to generate a random sample in the reverse process, the latent $x_{_T, g_m}$ needs to be roughly an isotropic Gaussian distribution. In other words, the relevant variables including $\alpha_t$ must be very close to zero and $\beta_{t}$  should also have a small value to force the $x_{_T, g_m} \sim \mathcal{N}(0,I)$. 
% \textbf{XXX should I discuss scheduler or not and cite Ho et al.? XXX}
% It is important to note that the amount of noise added to the image every step is regulated by a schedule, which scales the mean and the variance. This ensures that the variance does not explode as more noise is added. For the noise schedule, we use the linear schedule implemented by Ho et al.\cite{ho-diffusion}.
\let\subsectionautorefname\sectionautorefname
\let\subsubsectionautorefname\sectionautorefname
The diffusion probabilistic model can be viewed similar to variational auto-encoders (VAE)\cite{p2-weighting}, where the reverse process $p_{\theta}$ is learned by a neural network( \autoref{sec:model-presentation}) and is equivalent to the decoder network in VAE. Contrary to VAE, the encoder in the diffusion model is a fixed forward diffusion process.

In the reverse process, our neural network $\epsilon_\theta$ with parameters of $\theta$ learns to denoise the given $x_{_t, g_m}$ and output the $x_{_{t-1}, g_m}$. With iterative subtraction of the noise predicted by the neural network ($\epsilon_{\theta}$), and starting with $x_{_{t-1}, g_m}$ which have standard Gaussian distribution, $x_{_T, g_m}$ can be written as\cite{ho-diffusion}:
\begin{equation}
    x_{_{t-1}, g_m} = \mathbb{C}_1 \left (           
    x_{_{t}, g_m} - \mathbb{C}_2\epsilon_{\theta}(x_{_{t}, g_m}, t, g_m)
    \right ) + \sigma_t z,
\end{equation}
where:
\begin{equation}
    \mathbb{C}_1 = (\sqrt{1-\beta_t})^{-1}, 
    \mathbb{C}_2 = \beta_t(\sqrt{1-\alpha_t})^{-1}, 
    \beta_t = {\alpha_t}^2
\end{equation}
Similarly, $x_{_0, g_m}$ which is the generated image at the end of iterations, can be written as:
\begin{equation}
    x_{_0, g_m} = \frac{1}{\sqrt{1-\beta_1}} \left (           
    x_{_{1}, g_m} - \frac{\beta_1}{\sqrt{1-\alpha_1}}\epsilon_{\theta}(x_{_{1}, g_m}, 1, g_m)
    \right ) + \sigma_1 z,
\end{equation}
\subsubsection{Training Loss}
The final objective for training the utilized Diffusion probabilistic model is a combination of score matching losses~\cite{ScoreMatching} that can be summarized as the following:

\begin{equation}
    Loss = L_{simple} + cL_{vlb},
\end{equation}
where:
\begin{equation}
    L_{simple} = \sum_t \lambda_{t}L_{t}, \; \;
    L_{vlb} = \sum_t L_t,
\end{equation}
$L_t$ is a score matching loss for the time step $t$ which looks at the difference between the two Gaussian distributions. It can be written as:
\begin{equation}
\begin{split}
    L_t &= D_{KL}(q(x_{_{t-1},g_m}|x_{_t,g_m},x_{_0,g_m})\big|p_\theta(x_{_{t-1},g_m}|x_{_t,g_m})) \\ &=\mathbb{E}_{x_{_0, g_m},\epsilon}[\frac{\beta_t}{(1-\beta_t)(1-\alpha_t)}\|\epsilon - \epsilon_\theta(x_{_t,g_m},t)\|^2],
\end{split}
\end{equation}
$L_{simple}$ was initially proposed by Ho et al.~\cite{ho-diffusion} and use the following wights:
\begin{equation}
    \lambda_{t} = \frac{(1-\beta_t)(1-\alpha_t)}{\beta_t}.
\end{equation}
Considering the $\lambda_{t}$ values, $L_{simple}$ refers to a mean-squared error(MSE) loss defined on the difference of the actual and estimated noise, but Nicole et al.~\cite{diffusion-nichol} added the second term to the loss function to learn the $\sigma_t$ and showed that a small value for $c$ can significantly improve the model's capacity.
\subsubsection{Morphology Levels Prioritization}
% \textbf{Move it to method}
% More recently, Choi et al.\cite{p2-weighting} introduced perception prioriterized (P2), which aims to prioriterize solving the pretext task of more important noise levels. Higher weights are assigned to the loss at levels where the model learns perceptually rich contents while minimal weights are assigned to the loss at levels with imperceptibe details. 
% This new design significantly improved the performance of diffusion probabilistic models.
\textit{Signal-to-noise-ratio} (SNR) of the noisy image at the time step $t$ ($x_{_t,g_m}$) based on \autoref{eq:eq12} is equivalent to the following:
\begin{equation}
    SNR(t) = \frac{\alpha_t}{1-\alpha_t},
\end{equation}
Given the diminishing nature of SNR(t), it is demonstrated that the model concentrates rough and coarse properties during the early phases of the reverse diffusion process (when SNR is lower). Then, in the middle steps, it focuses on the image's perceptual components, while the latter stages (with the highest SNR) are dedicated to imperceptible minutiae~\cite{p2-weighting}.
Histology images are fairly sensitive that requires more accurate features. Similarly, our model should be focused on learning pathological and morphological markers that pathologists need to make a diagnosis at intermediate steps before performing minor denoising tasks at the end.
For morphology prioritization, the $\lambda_t$ weights can be utilized to devote heavier weights to the loss at earlier levels to emphasise perceptual contents and lower weights to the later levels.
We observed empirically that perception prioritized weighting provided by Choi et al.~\cite{p2-weighting} can result in generating higher detailed histopathology images:
\begin{equation}
    \lambda'_t = \frac{\lambda_t}{(k+SNR(t))^\gamma},
\end{equation}
where $k$ and $\gamma$ are used to keep the $\lambda'_t$ from extraordinarily increasing for very low SNR values and to control the concentration on clean-up details, respectively.

% where $k$ determines sharpness of the weighting scheme and prevents exploding weights for extremely small SNRs, and $\gamma$ controls the strength of down-weighting focus on learning imperceptible details. 
% For the remainder of this paper, we refer to this weighting scheme as "P2". 
% Moreover, for $\gamma = 0$ this weighting scheme is the standard weighting scheme. 

\subsubsection{The Architecture}\label{sec:model-presentation}
  We chose the backbone neural network similar to the Unet based model improved by Dhariwal et al.~\cite{beat-gans}, which is inspired from the Unet model introduced by Ho et al.~\cite{ho-diffusion} for diffusion models. This model contains attention at three various resolutions that allows the model to concentrate on tiny features related to cells(e.g., cell shape or small blood veins) or larger elements like how cell distribution, the texture of the stroma or the overlaying tissue. It also benefits from BIGGAN downsampling/upsampling residual blocks\cite{BIGGAN} to maintain the model free of artefacts like checker boxes or aliasing, which may not be a vital issue for typical images but can completely disrupt the subtle and accurate patterns that should exist in histopathological images. It also uses embedding layer to inject timestep to the neural network. The rest of the weights of the model are shared between all the time steps. Moreover, genotypes are given to the model with a separated embedding layer similar to timesteps.

% The Unet architecture was introduced for the diffusion probabilistic model by Ho et al. \cite{ho-diffusion}. The Unet architecture consist of a stack of residual layers with downsampling convolutions followed by a stack of residual layers with upsampling convolutions. 
% Layers with the same spatial size are connected by skip connections. Moreover, attention layers with 4 heads are used at certain resolutions. BigGAN residual blocks are used for upsampling and downsampling the activations. A class and timestep embedding are added and passed to the residuals blocks, in this way the model knows the timestep and class related to the sample. 
% For the rest of the architecture, we use 64 base channels, and 64 channels per head.
% Every diffusion probabilistic model used within this paper is trained with batch size 4, 1000 diffusion steps, learning rate $1e-4$, dropout $0.0$, and more than 150K iterations. And, for every 10K iterations the diffusion probabilistic model is saved.

\section{Experimental Evaluation and Results}\label{sec:results}
In this section, we assess the performance of the proposed approach for utilizing diffusion probabilistic models on generating synthetic histopathology images and compare it against one of the closest works, using our unique slide annotations collected by certified specialists. We also report the results of a survey in which two pathologists rated the quality of the generated images.

% \textbf{For each section of evaluation, describe that we used data like this to asses folan. and we use all the images and generated 50K to FId or sepaprate them to train and separate based on patients to keep from leakage and generate 54L images. }

% \section{Results and Discussion}

% \begin{table}[h]
% \centering
% %\setlength{\tabcolsep}{10pt}
% \setlength{\tabcolsep}{6pt}
% %\setlength{\tabcolsep}{2em}
% \renewcommand{\arraystretch}{1.0}
% \begingroup
% \footnotesize
% \tiny
% \begin{tabular}{llllll}
% \hline
%  & DPM1 & DPM2 & DPM3 & DPM4 & DPM5\\
% \hline
% Image size & 512 & 512 & 512 & 64 & 128\\
% Attention resolution & 32,16,8 & 32,16,8 & 64 & 32,16,8 & 32,16,8\\
% Weighting scheme & standard & P2 & P2 & P2 & P2\
% \end{tabular}
% \endgroup
% \caption{Hyperparameters for diffusion probabilistic models used in this paper.}
% \label{tab:model-hyperparameters}
% \end{table}

\subsection{Data}

\begin{figure*}[tp]
\rotatebox[origin=c]{90}{\bfseries ProGAN\strut}
\begin{subfigure}{.31\textwidth}
  \centering
  \caption{IDHC}
  % include first image
  \includegraphics[width=0.49\textwidth]{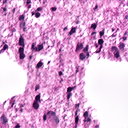} 
  % include second image
  \includegraphics[width=0.49\textwidth]{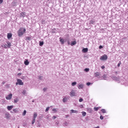}
\end{subfigure}
\begin{subfigure}{.31\textwidth}
  \centering
  \caption{IDHNC}
  % include third image
  \includegraphics[width=0.49\textwidth]{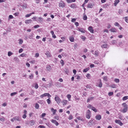} 
  % include fourth image
  \includegraphics[width=0.49\textwidth]{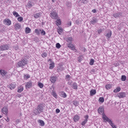}
\end{subfigure}
\begin{subfigure}{.31\textwidth}
  \centering
  \caption{IDHWT}
  % include fifth image
  \includegraphics[width=0.49\textwidth]{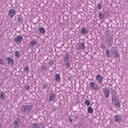} 
  % include sixth image
  \includegraphics[width=0.49\textwidth]{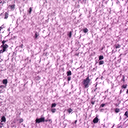}
\end{subfigure}
\newline
\newline
\rotatebox[origin=c]{90}{\bfseries Diffusion \strut}
\begin{subfigure}{.31\textwidth}
  \centering
  % include first image
  \includegraphics[width=0.49\textwidth]{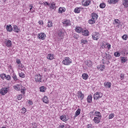} 
  % include second image
  \includegraphics[width=0.49\textwidth]{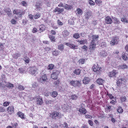}
\end{subfigure}
\begin{subfigure}{.31\textwidth}
  \centering
  % include third image
  \includegraphics[width=0.49\textwidth]{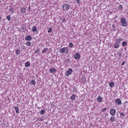} 
  % include fourth image
  \includegraphics[width=0.49\textwidth]{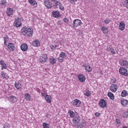}
\end{subfigure}
\begin{subfigure}{.31\textwidth}
  \centering
  % include fifth image
  \includegraphics[width=0.49\textwidth]{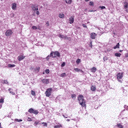} 
  % include sixth image
  \includegraphics[width=0.49\textwidth]{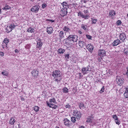}
\end{subfigure}
\caption{Selection of generated patches with diffusion and ProGAN models. First row represents synthesized images by ProGAN and second row refers images produced by diffusion model. Each dual columns together shows the generated samples representing one of the three different subtypes, namely IDHC, IDHNC, and IDHWT.}
\label{fig:good-generated-images}
\end{figure*}

We utilized a dataset of 344 whole slide images (WSIs) of low grade gliomas representative of its three major genomic subtypes from the Cancer Genome Atlas (TCGA) archive~\cite{dataset}. The dataset includes 297 cases with IDH mutations and 47 IDH Wild Type cases. The IDH Wild Type group has no IDH mutations and is labeled as IDHWT. Furthermore, the 297 IDH mutant slides are further divided into two groups: with no 1p19q chromosomal codeletion (173 slides) and with 1p19q codeletion (124 slides) labeled as IDHNC and IDHC, respectively. Each WSI is a large scale image with the size of $\sim100K\times100K$ pixels. Moreover, mutations associated with each patient were obtained from cbioportal (https://www.cbioportal.org/).

Each slide is pixel-wise annotated with an emphasis on the tumor-rich areas and attempted to avoid artefacts and empty spaces by a board-certified pathologist or a pathology resident under the supervision of a board-certified pathologist using our online annotation tool. These annotations will be made available to other researchers.

Annotated tumor areas from each slide were divided into small image tiles (referred to as patches) at specified objective magnification levels to improve computing performance.
% For computational efficiency, annotated tumor regions from each WSI were tiled to image patches in desired optical magnification levels. 
A maximum of 100 512$\times$512 pixel patches from the tumor annotated regions were taken from each slide at original magnification of 40$\times$ with a stride of 512 and scaled to 128$\times$128 patches, resulting in a final magnification of 10$\times$. The pixel size at full-resolution was $\sim0.25\mu$m and down sampled to $\sim1\mu$m. Finally, a total of 33,777  128$\times$128 pixel patches (at 10$\times$ magnification) were extracted from the WSIs and used to train various conditional diffusion probabilistic models. \autoref{tab:number-of-patches} provides the breakdown of the extracted patches based on genomic subtypes (Figure 1 in supplementary material shows examples  of extracted patches). 
% Worth noting that we will publicly release our code, selected patients, and their slides annotations for other researchers, which can be used for training both patch level generative and discriminative histology models

\begin{table}[h]
\centering
\setlength{\tabcolsep}{6pt}
\renewcommand{\arraystretch}{1.0}
%\tiny
\scalebox{0.95}{
\begin{tabular}{l|llll}
 & IDHC & IDHNC & IDHWT & Total \\
\hline
Patches & 12,139 & 16,975 & 4,663 & 33,777\\
IDH Status & Mutant & Mutant & Wildtype & - \\
1p19q Status & Codeletion & Retained & - & - \\
\end{tabular}}
\caption{Breakdown of extracted patches per subtype}
\label{tab:number-of-patches}
\end{table}

% \subsection{Model Implementation and Training} 
% We trained our model using PyTorch with 1000 diffusion steps using a workstation with an NVIDIA V100 GPU. The rest of the parameters are available at~\autoref{tab:model-hyperparameters}.
% \textbf{XXXX should we cite the github implementation of the paper that we used??? XXXX}

% \begin{table}[h]
% \centering
% \setlength{\tabcolsep}{10pt}
% %\setlength{\tabcolsep}{7pt}
% %\setlength{\tabcolsep}{2em}
% \renewcommand{\arraystretch}{1.0}
% %\tiny
% \begin{tabular}{l|l}
% \hline
% Parameter &  Value\\
% \hline
% Image size &  128\\
% Weighting scheme &  P2\\
% Diffusion steps & 1000\\
% Noise schedule & linear\\
% Maximum Patches & 100\\
% Channels &  64\\
% Heads &  4\\
% Heads channels & 64\\
% Attention resolution  & 32,16,8\\
% BigGAN up/downsample & yes\\
% Num Resblocks & 2\\
% Dropout &  0.0\\
% Batch size  & 4\\
% Learning rate & 1e-4\\

% \end{tabular}
% \caption{Hyperparameters for diffusion models used in this paper.}
% \label{tab:model-hyperparameters}
% \end{table}

\subsection{Experiments}

We evaluate the model's performance in different unique scenarios to thoroughly examine the various objectives that the model should achieve. We also utilize several objective metrics to assess the quality of generated images based on each experiment's specific requirements. Model implementation and training details are available at Table 1, section 1.2 in the supplementary material.

\subsubsection{Experiment I}
The objective of this experiment is to compare and contrast the quality of the synthesized images by our diffusion probabilistic model against a state-of-the-art study in which Levine et al. ~\cite{paper-ali} utilized ProGAN~\cite{ProGAN}. In their study, the authors showed the superiority of histology images generated by ProGAN relative to other generative models such as variational autoencoder~\cite{vae}, enhanced super resolution GAN (ESRGAN)~\cite{ESRGAN}, and deep texture synthesis~\cite{DeepSyn}.

%The objective of this scenario is to evaluate the synthesized images by our diffusion probabilistic model against one of the latest GAN models proposed by Levine et al.~\cite{paper-ali} in 2021 for synthesizing histopathology images. They proposed using ProGAN~\cite{ProGAN} for this task and showed the superiority of histology images generated by ProGAN relative to some other generative models such as variational autoencoder~\cite{vae}, enhanced super resolution GAN (ESRGAN)~\cite{ESRGAN}, and deep texture synthesis~\cite{DeepSyn}. 
For a fair comparison, we utilized similarly normalized patches and due to the nature of our problem, we, inspired by~\cite{CGAN}, slightly modified ProGAN to generate histology images conditioned on genotypes. We also trained both models on all the available extracted patches. We present samples of synthetic images generated by both models in~\autoref{fig:good-generated-images} (more images at Figure 2 in supplementary material). These samples demonstrate higher quality of the images synthesized by our model compared with those generated by ProGAN.
% Two samples associated with each genotypes are shown in ~\autoref{fig:good-generated-images}.
Next, we compared the two models by randomly generating 50,000 images by each model and calculating two sets of metrics:

\textbf{1. Common Generative Evaluation Metrics:}
Three of the most widely used metrics for assessment of the generated images are: Inception Score (IS), Fréchet Inception Distance (FID), and sFID. We briefly discuss them in the following:

\paragraph{Inception Score (IS):}
We report Inception Score~\cite{IS}, which is defined as:
\begin{equation}
    IS = \exp \left [ E_{x\sim{p_g}} D_{KL} \left ( p(y|x) || p(y)\right )  \right ],
\end{equation}
where $p(y)$ is marginal class probability and $D_{KL}$ is the KL-divergence.
\paragraph{Fréchet inception distance (FID):}
This metric compares the distribution of generated images with the real images' distribution in Inception-V3 latent space\cite{FID}. The more similar the synthetic images are to the input patches, the lower value that FID will have. The real and synthetic data are fed into the inception V3 model, and FID compares the mean and the standard deviation of the features extracted from pool\verb|_|3 layer. The FID is given by:
\begin{equation}
    FID (\mu_{r},\Sigma_{r},\mu_{g},\Sigma_{g}) = 	\|\mu_{r}-\mu_{g}\|^2 + {T_r}(\Sigma_{r}+\Sigma_{g}-2(\Sigma_{r}\Sigma_{g})^{\frac{1}{2}}),
\end{equation}
where $\mu_{r}$ and $\mu_{g}$ are the mean of the real and synthetic samples' embeddings. Similarly, $\Sigma_{r}$ and $\Sigma_{g}$ refer to their covariance.

% ($\mu_{r}, \Sigma_{r}$) and ($\mu_{g}, \Sigma_{g}$) as the sample mean and covariance of the embeddings of the real and synthetic data, respectively.

\paragraph{sFID:}
This is a modified version of FID proposed by Szegedy et al.~\cite{sFID} that uses the initial channels from an intermediate layer to compare the means and standard deviations.

IS may not be a suitable statistic for generative models trained on datasets other than ImageNet, as noted by Barratt et al.~\cite{WhyISno}. As a result, evaluating the model using FID and sFID is critical. FID is less sensitive to spatial heterogeneity since it is calculated using features from one of the latest layers that compresses spatial information. However, sFID employs intermediate features, which can detect spatial similarity better than FID in some situations~\cite{sFIDvsFID}. Reporting these together can measure the quality of the generated samples, which are summarized in \autoref{tab:Exp1_set1}. The results indicate that the proposed diffusion model outperforms the state-of-the-art across all these metrics. Also, lower values of both FID and sFID with extracted features from different layers make them sensitive to small changes and is able to detect mode coverage. This shows that unlike ProGAN the diffusion model is capable of producing perceptual features robustly.

\begin{table}[h]
\centering
% \setlength{\tabcolsep}{8pt}
%\setlength{\tabcolsep}{7pt}
%\setlength{\tabcolsep}{2em}
% \renewcommand{\arraystretch}{.9}
%\tiny
\scalebox{0.99}{
\begin{tabular}{l|l|l}
% \hline
 &  ProGAN & Diffusion Model\\
%  & & (ours)\\
\hline
Inception Score & 1.67 & \textbf{2.08} \\
\hline
FID & 53.85 & \textbf{20.11}\\
\hline
sFID & 24.37 & \textbf{6.32}\\
% \hline

\end{tabular}}
\caption{Summary of the Inception Score, FID, ans sFID for the first experiment}
\label{tab:Exp1_set1}
\end{table}

% image for distribution of manifolds of that weird paper

\begin{figure}[tp]
    \includegraphics[scale=0.4]{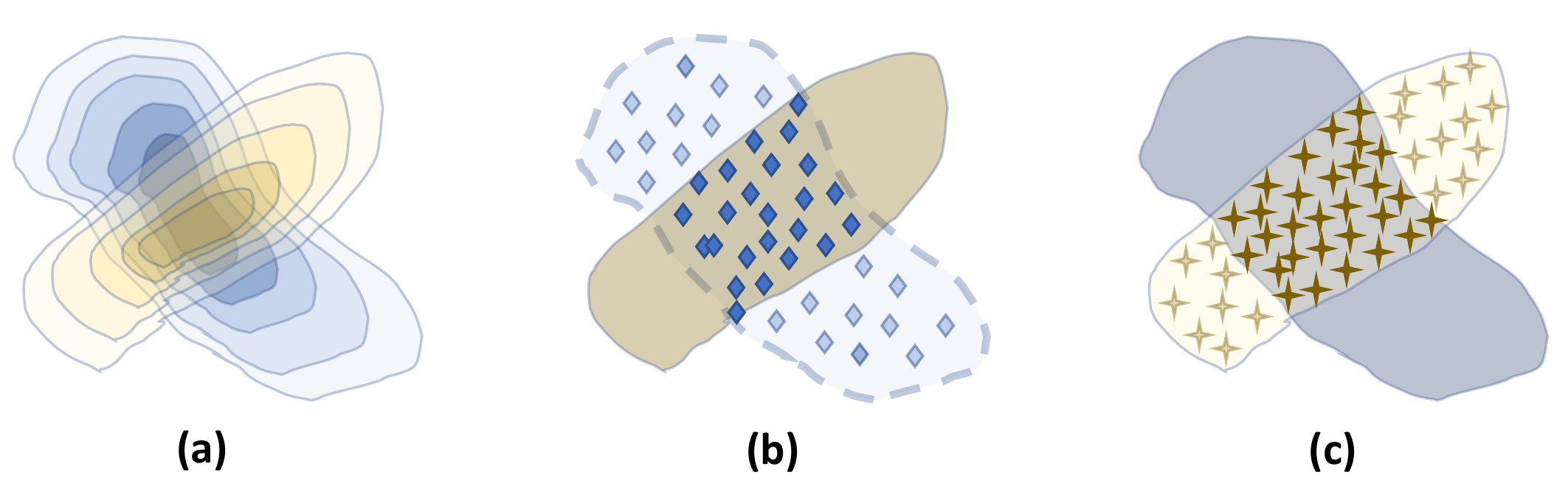}
    \caption{Real and synthetic data are depicted by blue and brown colors, respectively. (a) Presents the manifold for real and generated distributions. (b), and (c) Illustrate the Improved Recall and Precision, respectively.}
    \label{fig:HistoProcess}
\end{figure}

\textbf{2. Improved Precision and Recall Metrics:}
Kynk{\"a}{\"a}nniemi et al.~\cite{ImprovedRecall} discussed that both the quality and distribution coverage of the produced samples are essential for evaluating generative models. The authors proposed two metrics; namely "Improved Recall" and "Improved Precision" that 
can estimate both attributes by constructing non-parametric approximations of real and synthetic domain manifolds. 
We begin by estimating feature manifolds by computing distances to k-NN for each sample. Following that, "Improved Precision" refers to the percentage of produced samples inside the actual data manifold, while "Improved Recall" refers to the ratio of real samples located in the synthetic manifold.
% can estimate both aspects by forming explicit and non-parametric representations of the manifolds of real and synthetic data. We first estimate manifolds of features by calculating distances to k-NN of each sample. Next, the ratio of generated samples within the real data manifold is called "Improved Precision" and the ratio of real samples located in the synthetic manifold is called "Improved Recall". 

These two concepts are depicted in~\autoref{fig:HistoProcess}, and the results are summarized in~\autoref{tab:Exp1_set2}. We can conclude that the proposed method produces better images than the state-of-the-art in terms of both diversity and fidelity. Also, it shows that our model is able to significantly differentiate morphological features of histology images.
\begin{table}[h]
\centering
% \setlength{\tabcolsep}{8pt}
%\setlength{\tabcolsep}{7pt}
%\setlength{\tabcolsep}{2em}
% \renewcommand{\arraystretch}{0.99}
%\tiny
\scalebox{0.99}{
\begin{tabular}{l|l|l}
% \hline
 &  ProGAN & Diffusion Model\\
%  & & (ours)\\
\hline
Improved Recall & 0.4816 & \textbf{0.8528} \\
\hline
Improved Precision & 0.0078 & \textbf{0.2573}\\
% \hline
\end{tabular}}
\caption{Summary of the Improved Recall and Precision for the first experiment}
\label{tab:Exp1_set2}
\end{table}

\subsubsection{Experiment II}
The purpose of this experiment is to compare the morphological properties of synthetic and actual images. 
We selected an equal number of real and synthetic images generated by our diffusion model and designed a pathologist survey consisting of the following two questions. 
The first question asks if the participants believe the image is real or synthetic, and the second question inquires about their confidence level (more details on the survey are available at Figure 3 and Figure 4, Section 1.5 in supplementary material). 
The images were displayed according to a random order. 
Two pathologists participated with varying levels of expertise participated in this survey: a board-certified pathologist (P1) and a pathology resident (P2). 
The summary of the results is given in \autoref{tab:Exp2}, which shows that all participating experts could not distinguish the real  from synthetic images generated by our diffusion model. 
For the majority of small percentage of synthetic images that experts were able to correctly identify, they indicated less confidence level. 
Our survey results show that our synthetic histopathology images look extremely similar to real examples, making them an excellent candidate for a variety of real-world applications.

We also utilized two sided Fisher-exact test to examine whether there is a statistically significant difference between each pathologist observations for the real and synthetic images (p-values are available in \autoref{tab:fisher_test}). The resulting p-values demonstrate there is no statistically significant difference between performance of the pathologists on identifying real versus synthetic images.
% and shows that based on pathologists opinion, they have the same distribution.

\begin{table}[h]
\centering
\caption{Summary of results for Exp. II}
    \begin{subtable}{0.5\textwidth}
    \centering
    
    \scalebox{0.9}{
    \begin{tabular}{l|l|l|l|l||l|l}
    % \hline
    &  \textbf{Real}  & \textbf{Real}  &  \textbf{Syn.}  & \textbf{Syn.}&  \textbf{Real}  & \textbf{Syn.}\\
    % \hline
    \textbf{Conf.} &  High  & Med.  &  Med.  & High &  All  & All\\
    \hline
    \textbf{Real GT} & 0.75 &	0.05 & 0.175 & 0.025 & \textbf{0.8}	& \textbf{0.2}\\
    \hline
    \textbf{Syn. GT} &  0.775 & 0.05 & 0.125 & 0.05 & \textbf{0.825} & \textbf{0.175}\\
    % \hline
    
    \end{tabular}}
    \label{tab:P1}
    \caption{Summary of the results for P1}
    \end{subtable}
    
    \begin{subtable}{0.5\textwidth}
    \centering
    
    \scalebox{0.9}{
    \begin{tabular}{l|l|l|l|l||l|l}
    % \hline
    &  \textbf{Real}  & \textbf{Real}  &  \textbf{Syn.}  & \textbf{Syn.}&  \textbf{Real}  & \textbf{Syn.}\\
    % \hline
    \textbf{Conf.} &  High  & Med.  &  Med.  & High &  All  & All\\
    \hline
    \textbf{Real GT} & 0.225 & 0.2 & 0.25 & 0.325 & \textbf{0.425}	& \textbf{0.575}\\
    \hline
    \textbf{Syn. GT} &  0.325 & 0.25 & 0.2	& 0.225 & \textbf{0.575} & \textbf{0.425}\\
    % \hline
    
    \end{tabular}}
    \caption{Summary of the results for P2}
    \label{tab:P2}
    \end{subtable}
    % \begin{subtable}{0.5\textwidth}
    % \centering
    % \scalebox{0.85}{
    % \begin{tabular}{l|l|l|l|l||l|l}
    % % \hline
    % &  \textbf{Real}  & \textbf{Real}  &  \textbf{Syn.}  & \textbf{Syn.}&  \textbf{Real}  & \textbf{Syn.}\\
    % % \hline
    % \textbf{Conf.} &  High  & Med.  &  Med.  & High &  All  & All\\
    % \hline
    % \textbf{Real GT} & 0 & 0.45 & 0.55 & 0 & \textbf{0.45}	& \textbf{0.55}\\
    % \hline
    % \textbf{Syn. GT} &  0 & 0.825 & 0.175 & 0 & \textbf{0.825} & \textbf{0.175}\\
    % % \hline
    % \end{tabular}}
    % \caption{Summary of the results for P3}
    % \label{tab:P3}
    % \end{subtable}
    \label{tab:Exp2}
\end{table}

\subsection{Visual Observation:}
\begin{figure}[tp]
\begin{center}
    \includegraphics[scale=.84]{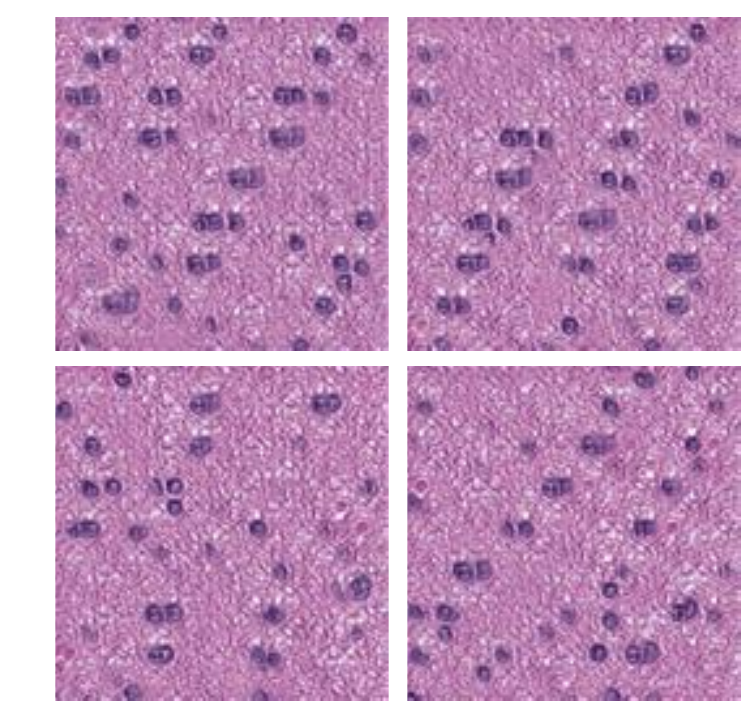}
    \caption{Examples of failed cases generated by ProGAN }
    \label{fig:ModeCollapse}
\end{center}
\end{figure}

\begin{table}[h]
    \centering
    \scalebox{1.0}{
    \begin{tabular}{l|l|l}
    &  \textbf{P1}  & \textbf{P2} \\
    \hline
    \textbf{Fisher-exact's p-value} & 1.0 & 0.26347\\
    
    \end{tabular}}
    \caption{Summary of the fisher test results}
    \label{tab:fisher_test}
\end{table}

The bottom row in \autoref{fig:good-generated-images} shows the images generated by our diffusion model. In these images, cell nuclei stained by purplish blue and extracellular matrix and cytoplasm stained by pink color related to H$\&$E staining, which suggests that using the color normalization module is effective.
Although medical inspection of generated images should be done by pathologists, there are some well known histology features in LGG images that can be identified even by an untrained person. The so-called "fried egg" appearance~\cite{OligoPattern} of oligodendrogliomas is shown in the synthetic images (such as the bottom first and second images in~\autoref{fig:good-generated-images}, top second image or bottom fourth image at Figure 2 in the supplementary material), namely in the IDHC and IDHWT. Another characteristic of the oligodendrogliomas that can be seen in generated images by diffusion (such as the bottom second column of~\autoref{fig:good-generated-images}) is branching small, chicken wire-like blood vessels~\cite{OligoPattern}, and this characteristic can also be found in the IDHC. This suggests that the diffusion model was able to learn specific known histopahological features. However, such specific features do not clearly exist in the images generated by ProGAN. In addition, the IDHWT are the most uncommon among the other two, and it appears that the lesser number of cases for the IDHWT subtype resulted in artefacts and lower image quality of ProGAN as compared to the diffusion, implying a mode collapse in ProGAN (\autoref{fig:ModeCollapse}) due to a lack of enough data points in this subtype. However, the diffusion model could learn these rare class-specific features.  \autoref{fig:ModeCollapse} shows samples of failed images by ProGAN in which each image is produced from random noise; however, they are fairly similar to each other.

\section{Conclusion and Future Work}

% Cite Vahdat work
% Results from P2 and P3 show the importance of a using synthetic images in pathologists training process to help them have similar expertise as P1. 
%To the best of our knowledge, this is the first work that utilizes diffusion probabilistic models for synthesizing histopathology images. 
We proposed an end-to-end method based on diffusion probabilistic models to generate H\&E stained histopathology images. 
To our knowledge, this is the first work that utilizes such models for histopathology image synthesis.  
%To train our model, we utilized pixel-wise annotations on a large cohort of digitized H\&E slides from TCGA LGG cohort. 
Using multiple objective and subjective metrics, we compared the performance of our proposed approach to proGAN, that has shown remarkable performance in generating histopathology images. 
Results suggest that our proposed approach outperforms proGAN. 
Additionally, we conducted an empirical study where pathologists participated in a survey in which they were not able to distinguish the synthetic from real images.  
Taken together, the proposed method could facilitate the deployment of synthesized histology images for many real-life educational, privacy, and data augmentation applications. 

In addition, the work in this paper can be extended by optimising the proposed model to reduce the sampling time of the diffusion probabilistic models, which is relatively longer than GANs due to multiple small diffusion steps. As an instance, this work can be expanded by drawing on the work of Xiao et al.~\cite{vahdat-problem}, who use a multimodal conditioned discriminator that follows the diffusion model and can significantly reduce the number of diffusion steps.

{\small
\bibliographystyle{ieee_fullname}
\bibliography{egbib}
}

\newpage
\onecolumn

\section{Supplementary Materials}\label{sec:supplementary}

This section provides additional information that we could not include in the paper itself because of space limitations.

\subsection{Forward Diffusion in Detail}\label{sec:supp_forward}
This diffusion process $q$ is a Markov chain and produces latents $x_{_1, g_m}$ through $x_{_T, {g_m}}$ as follows~\cite{ho-diffusion}:
\begin{equation}
    q(x_{_1, g_m}, ...., x_{_T, g_m}|x_{_0, g_m}) = \prod_{t=1}^{T} q(x_{_t, g_m}|x_{{_{t-1}, g_m}}),
\end{equation}
\begin{equation}
    q(x_{_t, g_m}|x_{_{t-1}, g_m}) = \mathcal{N}\left (x_{_t, g_m};\left (\sqrt{1-\beta_{t}} \right) x_{_{t-1}, g_m},\beta_{t}\mathit{I}\right ),
\end{equation}
and by using reparameterization, latent $x_{_t,{g_m}}$ can be derived directly from $x_{_0, g_m}$ as below:
\begin{equation}
    x_{_t,g_m} = \sqrt{\alpha_{t}}x_{_0,g_m}+\sqrt{1-\alpha_{t}}\epsilon,
\end{equation}
$0<\beta_{1},\beta_{2},...,\beta_{T}<1$ are fixed noise scales for each time step $t$ and $\alpha_t := \prod_{s=1}^{t}(1-\beta_s)$. Also, the distribution of the$\epsilon$ is as  $\epsilon \sim \mathcal{N}(0,I)$

\subsection{Model Implementation and Training}\label{sec:implementation}
We trained our model using PyTorch with 1000 diffusion steps using a workstation with an NVIDIA V100 GPU. The rest of the parameters are available at~\autoref{tab:model-hyperparameters}.

\begin{table}[h]
\centering
\setlength{\tabcolsep}{10pt}
\renewcommand{\arraystretch}{1.0}
%\tiny
\begin{tabular}{l|l}
\hline
Parameter &  Value\\
\hline
Image size &  128\\
Weighting scheme &  P2\\
Diffusion steps & 1000\\
% Noise schedule & linear\\
Maximum Patches & 100\\
Channels &  64\\
Heads &  4\\
Heads channels & 64\\
Attention resolution  & 32,16,8\\
BigGAN up/downsample & yes\\
Num Resblocks & 2\\
% Batch size  & 4\\
Learning rate & 1e-4\\

\end{tabular}
\caption{Hyperparameters for diffusion models used in this paper.}
\label{tab:model-hyperparameters}
\end{table}

\newpage

\subsection{Samples of Real Patches} 

\begin{figure*}[h]
\begin{center}

\rotatebox[origin=c]{90}{\bfseries IDHC\strut}
\begin{subfigure}{.145\textwidth}
  \centering
  % include first image
  \includegraphics[width=\textwidth]{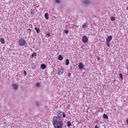} 
\end{subfigure}
\begin{subfigure}{.145\textwidth}
  \centering
  % include second image
  \includegraphics[width=\textwidth]{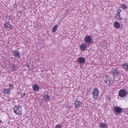}
\end{subfigure}
\begin{subfigure}{.145\textwidth}
  \centering
  % include third image
  \includegraphics[width=\textwidth]{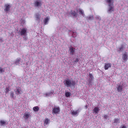}  
\end{subfigure}
\end{center}
\begin{center}
% \newline
\rotatebox[origin=c]{90}{\bfseries IDHNC\strut}
\begin{subfigure}{.145\textwidth}
  \centering
  % include first image
  \includegraphics[width=\textwidth]{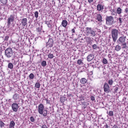} 
\end{subfigure}
\begin{subfigure}{.145\textwidth}
  \centering
  % include second image
  \includegraphics[width=\textwidth]{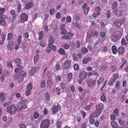}
\end{subfigure}
\begin{subfigure}{.145\textwidth}
  \centering
  % include third image
  \includegraphics[width=\textwidth]{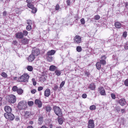}  
\end{subfigure}

\end{center}
% \newline
\begin{center}
\rotatebox[origin=c]{90}{\bfseries IDHWT\strut}
\begin{subfigure}{.145\textwidth}
  \centering
  % include first image
  \includegraphics[width=\textwidth]{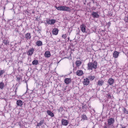} 
\end{subfigure}
\begin{subfigure}{.145\textwidth}
  \centering
  % include second image
  \includegraphics[width=\textwidth]{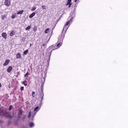}
\end{subfigure}
\begin{subfigure}{.145\textwidth}
  \centering
  % include third image
  \includegraphics[width=\textwidth]{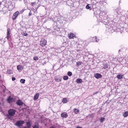}  
\end{subfigure}

\caption{Selection of patches with size =128x128 pixels extracted from the WSI. Each row represents a different genotypes.}
\label{fig:extracted}
    
\end{center}
\end{figure*}

\autoref{fig:extracted} shows sample of real extracted patches after applying the the color normalization module to them.

\subsection{Samples of Generated Images by Our Diffusion Probabilistic Model} 

\begin{figure}[h]
\begin{center}

    \includegraphics[scale=0.8]{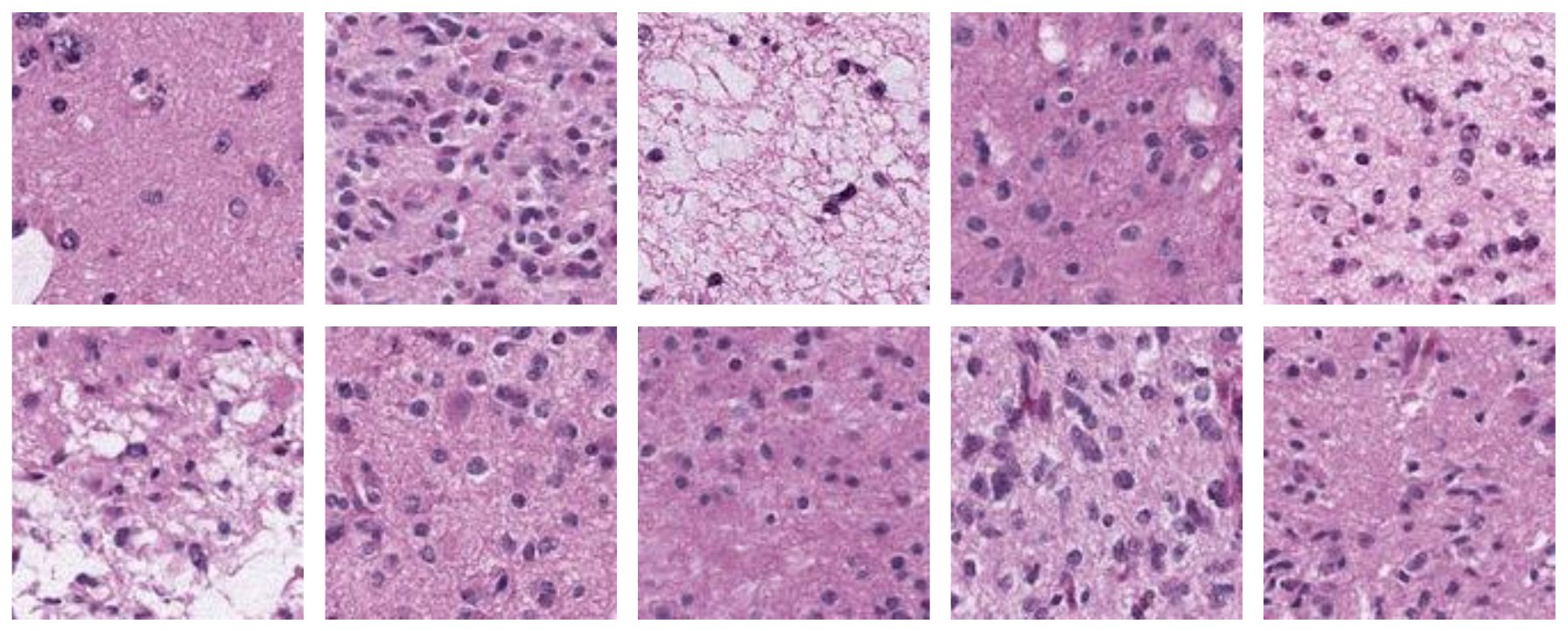}
    \caption{Top 10 generated images by our diffusion model.}
    \label{fig:Top10}
    
\end{center}
\end{figure}

\autoref{fig:Top10} presents the top 10 synthetic images produced by diffusion model.%, which at least two of pathologists classified them as real images with high confidence.

\newpage
\subsection{Survey Visualization}\label{sup:survey}
\begin{figure}[h]
\begin{center}

    \includegraphics[scale=1]{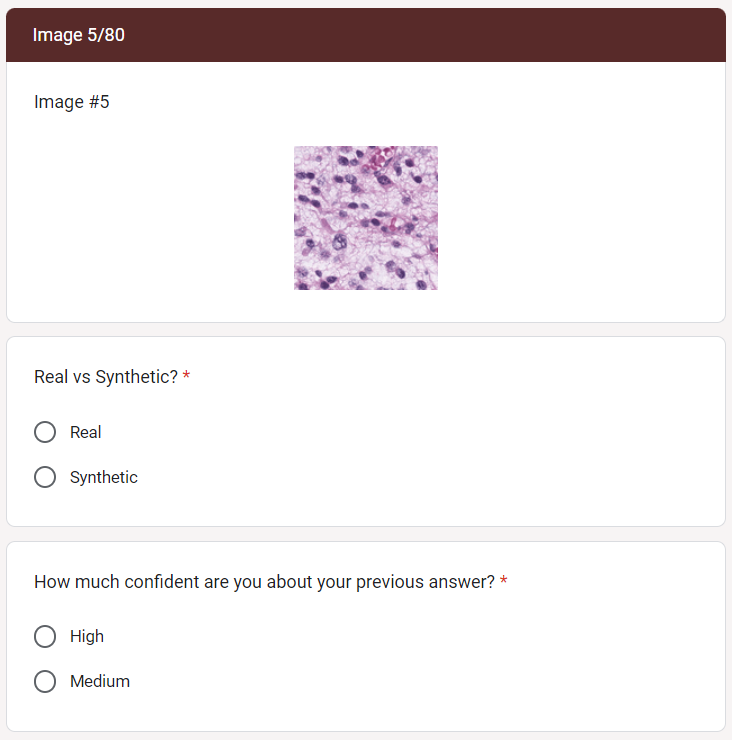}
    \caption{ The web form for subjective study}
    \label{fig:Form}
    
\end{center}
\end{figure}

\begin{figure}[h]
\begin{center}

    \includegraphics[scale=0.54]{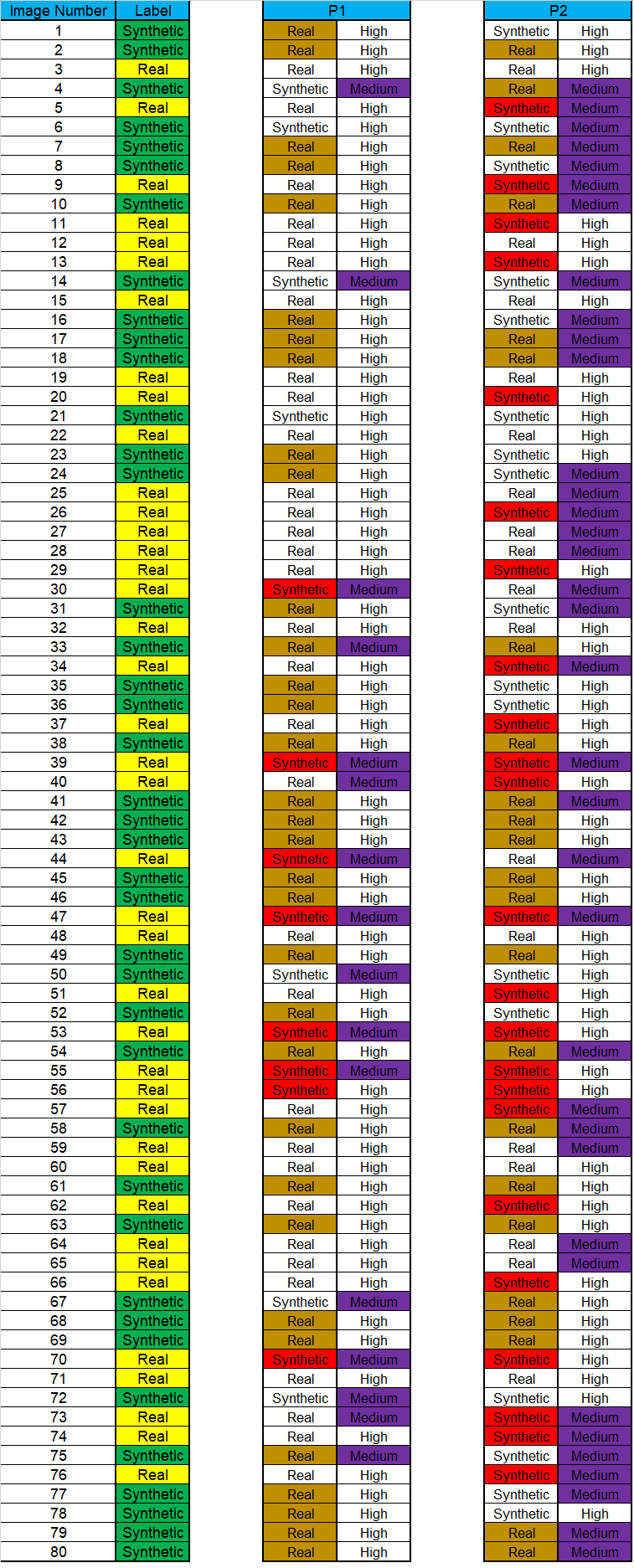}
    \caption{Illustration of each pathologist's opinion. Green: synthetic image, yellow: real one. Brown: synthetic images that classified as real by pathologists. Red: real image diagnosed as synthetic image. Purple: medium confidant answers.}
    \label{fig:Survey}
    
\end{center}
\end{figure}

\autoref{fig:Form} shows a sample section of the web-form that the pathologists completed. \autoref{fig:Survey} is the detailed illustration of each expert opinion about each real or synthetic image. Green and yellow colours depict synthetic and real images, respectively. Also, brown presents a synthetic images that classified as real by pathologists and red corresponds to a real images identified as synthetic. Purple shows medium confidence answers.

\end{document}